\documentstyle[12pt]{article}
\include{figures}
\begin{document}

\newcommand{\vslash}{\mbox{$\displaystyle\not\mkern-4mu v$}}

\mbox{ }\hfill{\normalsize UCT-TP 188/92}\\
\mbox{ }\hfill{\normalsize MZ-TH/92-54}\\
\mbox{ }\hfill{\normalsize November 1992}\\




\begin{center}

{\Large \bf Spectral Functions for Heavy-Light Currents\\[.5cm]
            and Form Factor Relations in HQET}
\vspace{1cm}

{\bf C.A. Dominguez}\footnote{Supported in part by the Foundation for
Research Development, South Africa.}\\[.5cm]
Institute of Theoretical Physics and Astrophysics\\
University of Cape Town, Rondebosch 7700, Cape, RSA\\[.3cm]
and\\[.3cm]
{\bf J.G. K\"{o}rner$^2$ and K. Schilcher}\footnote{Supported in part by
Bundesministerium f\"{u}r Forschung und Technologie BMFT, FRG under
$\; \; \; \;$ contract 06MZ730 and by the Humboldt Foundation.}\\[.5cm]
Institut f\"{u}r Physik, Johannes Gutenberg-Universit\"{a}t\\
Staudingerweg 7, D.6500 Mainz, Germany
\date{}
\vspace{.5cm}


\begin{abstract}


\noindent
We derive relations among form factors describing the current-induced
transitions:
(vacuum)  $\rightarrow B,B^{*}, B \pi, B^{*} \pi, B \rho$ and $B^{*}
\rho$ using heavy quark symmetry. The results are compared to
corresponding form factor relations following from identities between
scalar and axial vector, and pseudoscalar and vector spectral
functions in the heavy quark limit.
\end{abstract}
\end{center}

\newpage

With the introduction of the heavy quark effective theory
\cite{ref1},
($HQET$) we have witnessed dramatic developments in our understanding of
the physics of hadrons containing a heavy quark Q. The systematic
expansion in inverse powers of the heavy quark mass $m_{Q}$
accomplished in the $HQET$ allows QCD calculations of hadronic
processes to a level of rigor previously only conceivable in deep
inelastic reactions. In the heavy quark limit (HQL) the effective
Lagrangian exhibits a new spin-flavor symmetry \cite{ref1}. This
heavy quark symmetry (HQS) imposes restrictive constraints on weak
decay amplitudes. Notable results are the scaling relation between
decay constants \cite{ref2} and the reduction of semileptonic
form factors of heavy mesons and baryons to a small number of
Isgur--Wise functions \cite{ref1}.

In this paper we will present an alternative approach to obtain
relations between form factors in the heavy quark limit ($HQL$). Our
approach is based on the observation that in this limit certain
correlators of two currents comprised of a heavy and a light quark
become identical (while they bear no relation to each other in the
full QCD). For example, the vector-vector (VV) correlator equals the
pseudoscalar-pseudoscalar (PP) one and the axial vector-axial vector
(AA) correlator equals the scalar-scalar (SS) one. If we adopt the
point of view that the physical spectral function is obtained from
the QCD correlator by some form of analytic continuation, then
{\bf identical} QCD correlators imply {\bf identical}
physical spectral functions. We will try in this paper to exploit
this form of duality as far as possible.

We consider two-point functions defined (in full QCD) through
\begin{equation}
\Pi_{\mu \nu}(q) = i \int dx e^{iqx} \langle 0|T \left( J_{\mu} (x)
J_{\nu}^{\dagger} (0) \right)| 0 \rangle
\end{equation}
where the renormalized currents are
\begin{equation}
J_{\mu}(x) = \bar{q}(x) \Gamma_{\mu} Q(x) , \;\;\;\;\;
\Gamma_{\mu} = \gamma_{\mu} \;\; or \;\; \gamma_{\mu} \gamma_{5}
\end{equation}
with $q(x) (Q(x))$ being the light (heavy) quark field.

Given a hadron with four-velocity $v_{\mu}$ and assuming that the
four-velocity of the heavy quark is almost equal to $v_{\mu}$ one can
factorize out the large $m_{Q}$ effects by introducing a field
$h_{v}(x)$ through
\begin{equation}
h_{v}(x) = \frac{1 + \vslash}{2} e^{im_{Q}v \cdot x} Q(x)
\end{equation}
which carries only the residual momentum. When the QCD Lagrangian is
expressed in terms of the field $h_{v}(x)$ it exhibits the well-known
static spin flavour symmetries.

The current (2) then goes over to
\begin{equation}
J_{\mu}(x) \rightarrow e^{-im_{Q}v \cdot x} \tilde{J}_{\mu}(x)
\end{equation}
where $\tilde{J}_{\mu}(x)$ is the current in the effective theory.
\begin{equation}
\tilde{J}_{\mu}(x) = \bar{q}(x) \Gamma_{\mu} h_{v}(x)
\end{equation}
The matrix elements of $\tilde{J}_{\mu}$ satisfy the spin-flavor
symmetries. The $HQET$ predictions for decay parameters must, however,
still be related to the physical (i.e. full QCD) ones through the
procedure of running and matching.

The correlator in the effective theory reads
\begin{eqnarray}
\tilde{\Pi}_{\mu \nu}(q) &=& i \int dx e^{iqx} \langle 0 | T \left(
\tilde{J}_{\mu}(x) \tilde{J}_{\nu}^{\dagger}(0) \right) | 0 \rangle \nonumber
\\
                       &=& ( - g_{\mu \nu} q^{2} + q_{\mu} q_{\nu})
\tilde{\Pi}^{(1)} (q^{2}) + q_{\mu} q_{\nu} \tilde{\Pi}^{(0)} (q^{2})
\end{eqnarray}

In $HQET$ the following symmetry relations hold
\begin{eqnarray}
\tilde{\Pi}_{VV}^{(1)} &=& \tilde{\Pi}_{AA}^{(0)} \nonumber \\[1ex]
\tilde{\Pi}_{AA}^{(1)} &=& \tilde{\Pi}_{VV}^{(0)}
\end{eqnarray}
where we emphasize that the currents in the correlators (6) are
HQET currents.
The subscripts $VV$ and $AA$ refer to the currents involved in
the correlator (6).

The absorptive part of the two-point function is given by
\begin{eqnarray}
\mbox{Im} \; \; \tilde{\Pi}_{\mu \nu}(q) &=& (2 \pi)^{3} \int dx e^{iqx}
\sum_{n} \delta(q - p_{n}) \times \langle 0 | \tilde{J}_{\mu}(0) | n \rangle
\langle n |
    \tilde{J}^{\dagger}_{\nu}(0) | 0 \rangle \nonumber \\[1ex]
& & \equiv (- g_{\mu \nu} q^{2} + q_{\mu} q_{\nu}) \rho^{(1)}(q^{2}) +
q_{\mu} q_{\nu} \rho^{(0)} (q^{2})
\end{eqnarray}
Depending on the parity of the current $\tilde{J}_{\mu}$, different
intermediate states $n$ contribute to the spectral functions
$\rho^{(i)}
(i=0,1)$. The most important intermediate states which we will consider
in the following are listed in Table 1.
\begin{center}
\begin{tabular}{ll}
{\bf Table 1}: & Lowest lying intermediate states contributing to the\\
        & spectral functions $\rho_{VV}^{(i)}$ and $\rho_{AA}^{(i)}(i=0,1)$.
\end{tabular}\\[.5cm]
\end{center}
\begin{center}
\begin{tabular}{||c|l||} \hline \hline
Intermediate states & Spectral function  \\ \hline

$B$                 & $\rho_{AA}^{(0)}$  \\[1ex]
$B^{*}$             & $\rho_{VV}^{(1)}$  \\[1ex]
$B \pi$             & $\rho_{VV}^{(0)}$, $\; \;$  $\rho_{VV}^{(1)}$, $\; \;$
\\[1ex]
$B \rho$            & $\rho_{VV}^{(1)}$, $\; \;$  $\rho_{AA}^{(1)}$, $\; \;$
$\rho_{AA}^{(0)}$ \\[1ex]
$B^{*} \pi$         & $\rho_{VV}^{(1)}$, $\; \;$  $\rho_{AA}^{(1)}$, $\; \;$
$\rho_{AA}^{(0)}$ \\[1ex]
$B^{*} \rho$        & $\rho_{VV}^{(1)}$, $\; \;$  $\rho_{VV}^{(0)}$, $\; \;$
$\rho_{AA}^{(1)}$, $\; \;$ $\rho_{AA}^{(0)}$ \\[.3cm]
\hline \hline
\end{tabular}
\end{center}
In the $HQET$ \cite{ref3}
\begin{equation}
\rho_{VV}^{(1)} (q^{2}) = \rho_{AA}^{(0)} (q^{2})
\end{equation}
\begin{equation}
\rho_{AA}^{(1)} (q^{2}) = \rho_{VV}^{(0)} (q^{2})
\end{equation}
for $q^{2}$ close to the heavy-light threshold satisfying
$(q^{2} - m_{Q}^{2}) << m_{Q}^{2}$. From the identity Eq.(9) one
immediately concludes the equality of the leptonic decay constants
$f_{B}$ and
$f_{B^{*}}$ as a manifestation of the spin symmetry, where the decay
constants are defined as
\begin{eqnarray}
\langle 0|A_{\mu}|B \rangle &=& if_{B} p_{\mu} \nonumber  \\[1ex]
\langle 0|V_{\mu}|B^{*} \rangle &=& f_{B^{*}} M_{B^{*}} \epsilon_{\mu}
\end{eqnarray}
For the two particle intermediate states we expect relations between
various form factors. To determine these we first define the relevant
matrix elements,
\begin{eqnarray}
\langle B \pi|V_{\mu}|0 \rangle &=& \sqrt{M_{1}} (f_{1}^{V}v_{1 \mu}
- f_{2}^{V} v_{2 \mu}) \nonumber \\[1ex]
\langle B^{*} \pi|V_{\mu}|0 \rangle &=& \sqrt{M_{1}} f^{*V} i \epsilon
(\mu \epsilon_{1}^{*}v_{1} v_{2}) \nonumber \\[1ex]
\langle B^{*} \pi|A_{\mu}|0 \rangle &=& \sqrt{M_{1}} \epsilon_{1}^{* \alpha}
\left[ f_{1}^{* A} g_{\alpha \mu} + f_{2}^{* A} v_{2 \alpha} v_{1 \mu} +
f_{3}^{* A} v_{2 \alpha} v_{2 \mu} \right] \nonumber \\[1ex]
\langle B \rho |V_{\mu}|0 \rangle &=& \sqrt{M_{1}} g^{V} i \epsilon
(\mu \epsilon_{2}^{*} v_{1} v_{2}) \nonumber \\[1ex]
\langle B \rho |A_{\mu}|0 \rangle &=& \sqrt{M_{1}} \epsilon_{2}^{* \alpha}
\left[ g_{1}^{A} g_{\alpha \mu} + g_{2}^{A} v_{1 \alpha} v_{1 \mu} +
g_{3}^{A} v_{1_{\alpha}} v_{2_{\mu}} \right] \nonumber \\[1ex]
\langle B^{*} \rho |V_{\mu}|0 \rangle &=& \sqrt{M_{1}} \left[
\epsilon_{1 \alpha}^{*} \epsilon_{2 \beta}^{*}
v_{1 \beta} v_{2 \alpha} (v_{1 \mu} g_{1}^{* V} + v_{2 \mu}
g_{2}^{* V}) \right. \nonumber \\[1ex]
    & & +  \left. g_{\alpha \beta} (v_{1 \mu} g_{3}^{* V} + v_{2 \mu}
g_{4}^{* V}) + g_{\alpha \mu} v_{1 \beta} g_{5}^{* V} + g_{\beta \mu}
v_{2 \alpha} g_{6}^{* V} \right] \nonumber  \\[1ex]
\langle B^{*} \rho |A_{\mu}|0 \rangle &=& \sqrt{M_{1}} \left[
i \epsilon (\mu \epsilon_{1}^{*} \epsilon_{2}^{*}
v_{1}) g_{1}^{* A} + i \epsilon (\mu \epsilon_{1}^{*} \epsilon_{2}^{*}
v_{2}) g_{2}^{* A} \right. \nonumber \\[1ex]
    & &  + \left. i \epsilon (\mu \epsilon_{1}^{*} v_{1} v_{2})
         \epsilon_{2}^{*} v_{1} g_{3}^{* A} +
        i \epsilon (\mu \epsilon_{2}^{*} v_{1}v_{2}) \epsilon_{1}^{*}
        v_{2} g_{4}^{* A} \right]
\end{eqnarray}
where the indices 1 and 2 refer to the heavy and light meson,
respectively. The normalization of the heavy meson state is chosen to
be E/M to facilitate comparison with the $HQL$. We
use velocities $v_{i} = p_{i}/M_{i} (i = 1,2)$ rather than momenta in
the definition of the form factors so that they do not depend on the
heavy mass scale.

In the $HQET$ the form factor complexity is reduced. Denoting the
HQET heavy to light
reduced form factors by capital letters we obtain the following
results using the well-known trace formalism \cite{ref1} as
applied to heavy meson to light meson transitions (see also \cite{ref4})
\begin{eqnarray}
\langle B \pi | V_{\mu} | 0 \rangle &=& \sqrt{M_{1}}
(F_{1} v_{1 \mu} - F_{2} v_{2
\mu}) \nonumber \\[1ex]
\langle B^{*} \pi | V_{\mu} | 0 \rangle &=& \sqrt{M_{1}} (F_{2} i \epsilon (\mu
\epsilon_{1}^{*} v_{1} v_{2})) \nonumber \\[1ex]
\langle B^{*} \pi | A_{\mu} | 0 \rangle &=& \sqrt{M_{1}}
((F_{1} - \omega F_{2}) \epsilon_{1
\mu}^{*} + F_{2}(v_{2}\epsilon_{1}^{*}) v_{1 \mu}) \nonumber \\[1ex]
\langle B \rho | V_{\mu} | 0 \rangle &=& \sqrt{M_{1}}
(- G_{1} - G_{3}) i \epsilon (\mu \epsilon_{2}^{*} v_{1} v_{2}) \nonumber
\\[1ex]
\langle B \rho | A_{\mu} | 0 \rangle &=& \sqrt{M_{1}} \left\{
(\omega G_{1} - G_{2} - \omega G_{3} + G_{4}) \epsilon_{2 \mu}^{*}  - 2 G_{4}
v_{1} \epsilon_{2}^{*} v_{1 \mu} +
(- G_{1} + G_{3}) (v \cdot \epsilon_{2}^{*}) v_{2 \mu} \right\}
\nonumber \\[1ex]
\langle B^{*} \rho | V_{\mu} | 0 \rangle &=& \sqrt{M_{1}}
\left\{ - 2 G_{3} v_{1 \mu} (v_{1} \cdot \epsilon_{2}^{*})
(v_{2} \cdot \epsilon_{1}^{*}) + (G_{2} + 2 \omega G_{3} - G_{4})
(\epsilon_{2}^{*} \cdot \epsilon_{1}^{*}) v_{1 \mu} \right. \nonumber \\[1ex]
& &  \left. + (G_{1} + G_{3}) \left[ (v_{2} \cdot \epsilon_{1}^{*})
\epsilon_{2 \mu}^{*} - (\epsilon_{1}^{*} \cdot \epsilon_{2}^{*})
v_{2 \mu} \right] - (G_{2} + G_{4}) (v_{1} \cdot \epsilon_{1
\mu}^{*}) \right\} \nonumber \\[1ex]
\langle B^{*} \rho | A_{\mu} | 0 \rangle &=& \sqrt{M_{1}}
\left\{ ( - G_{2} - 2w G_{3} + G_{4}) i \epsilon (\mu \epsilon_{1}^{*}
\epsilon_{2}^{*} v_{1}) \right. \nonumber \\[1ex]
& & \left. + (G_{1} + G_{3}) i \epsilon (\mu \epsilon_{1}^{*}
\epsilon_{2}^{*} v_{2}) - 2 G_{3}(v_{1} \cdot \epsilon_{2}^{*}) i \epsilon
(\mu \epsilon_{1}^{*} v_{1} v_{2}) \right\}
\end{eqnarray}
where $\omega = (q^{2} - M_{1}^{2} - M_{2}^{2})/2 M_{1} M_{2}$. Some of these
results have been derived before \cite{ref4}. As a next task we would
like to demonstrate how Eqs.(12) and (13) follow from the equalities between
spectral functions Eqs.(9) and (10), or are consistent with them.

In order to be able to express our results in a compact form we will use
helicity form factors rather than the covariant ones of Eq.(12). In the
CM system with z-axis in the direction of the outgoing heavy particle
the polarization vectors $\epsilon_{\mu}(\lambda_{W})$ associated with the
currents are
\begin{eqnarray}
\epsilon_{\mu}(0) &=& (0,0,0,1), \; \; \; \epsilon_{\mu} (\pm) =
\frac{1}{\sqrt{2}}
                      (0, \mp 1, - i, 0) \nonumber \\[1ex]
\epsilon_{\mu}(t) &=& (1, 0, 0, 0)
\end{eqnarray}
where $\epsilon_{\mu}(0,\pm)$ and $\epsilon_{\mu}(t)$ refer to the spin 1
and spin 0 components of the currents, respectively.

We consider first the $B \pi$ intermediate states. The helicity form
factors are defined through\\[.3cm]
\begin{equation}
H_{{\lambda}_{W}}^{V} = \langle B \pi | V_{\mu} | 0
\rangle \cdot \epsilon^{\mu}(\lambda_{W}) \; .
\end{equation}

Using Eqs.(12) and (14) one obtains
\begin{eqnarray}
\sqrt{q^{2}}H_{0}^{V} &=& \sqrt{M_{1}} \sqrt{\omega^{2} - 1} \; \; \;
(- M_{2}f_{1}^{V} + M_{1}f_{2}^{V}) \nonumber \\[1ex]
\sqrt{q^{2}}H_{t}^{V} &=& \sqrt{M_{1}}\left( (M_{1} + M_{2} \omega)
f_{1}^{V} + (M_{2} + M_{1} \omega) f_{2}^{V} \right) \; .
\end{eqnarray}

In the $HQL$ ($M_{1} \rightarrow \infty$ with $\omega$ fixed, i.e
$\sqrt{q^{2}} \simeq M_{1}$) Eq.(16) reduces to\\[.3cm]
\begin{eqnarray}
H_{0}^{V} &=& - \sqrt{M_{1}} \sqrt{\omega^{2} - 1} \;
f_{2}^{V} \nonumber \\[1ex]
H_{t}^{V} &=& \sqrt{M_{1}} (f_{1}^{V} + \omega f_{2}^{V})
\end{eqnarray}
since $f_1^V$ and $f_2^V$ are heavy mass scale independent.

This result is of the same form as the HQET relation Eq.(13).

Next we turn to the $B^*\pi$ intermediate state.
The helicity amplitudes for the $B^{*} \pi$ intermediate state
are defined through
\begin{equation}
H_{\lambda_{W}=\lambda_{1}}^{A} = \langle B^{*}(\lambda_{1}),
\pi | A_{\mu}|0 \rangle \epsilon^{\mu}(\lambda_{W})
\end{equation}
In the $HQL$ one obtains
\begin{eqnarray}
H_{0}^{A} &=& \sqrt{\omega^{2}-1} \sqrt{M_{1}} \left[ - f_{1}^{*A} +
(\omega^{2} - 1) f_{3}^{*A} \right] \nonumber \\[1ex]
H_{t}^{A} &=& \sqrt{\omega^{2}-1} \sqrt{M_{1}} \left[ f_{2}^{*A}
+ \omega f_{3}^{*A} \right]   \nonumber \\[1ex]
H_{\pm}^{A} &=& - \sqrt{M_{1}} f_{1}^{*A} \nonumber \\[1ex]
H_{\pm}^{V} &=& \mp \sqrt{M_{1}} \sqrt{\omega^{2}-1} f^{*V} \; .
\end{eqnarray}

The condition $\rho_{VV}^{(0)} = \rho_{AA}^{(1)}$ now reads
\begin{equation}
3 |H_{t}^{V} (B \pi)|^{2} = |H_{0}^{A} (B^{*} \pi) |^{2} +
|H_{+}^{A}(B^{*} \pi) |^{2} + |H_{-}^{A} (B^{*} \pi) |^{2}
\end{equation}
or
\begin{equation}
3 | f_{1}^{V} + \omega f_{2}^{V} |^{2} = |f_{1}^{*A} -
(\omega^{2} - 1) f_{3}^{*A} |^{2} + 2 | f_{1}^{*A} |^{2} \; .
\end{equation}

If we assume pole behaviour of the helicity form factors a solution
of the equation (20) or (21) is
\begin{equation}
f_{3}^{*A} = 0, \; \; \; f_{1}^{*A} = f_{1}^{V} + \omega f_{2}^{V} \; .
\end{equation}

Similarly, one has from $\rho_{AA}^{(1)} = \rho_{VV}^{(0)}$
\begin{equation}
3 | H_{t}^{A} (B^{*} \pi) |^{2} = |H_{0}^{V}(B \pi) |^{2} +
  | H_{+}^{V}(B^{*} \pi) |^{2} + |H_{-}^{V}(B^{*} \pi) |^{2}
\end{equation}
or
\begin{equation}
3(\omega^{2} - 1) | f_{2}^{*A} - \omega f_{3}^{*A} |^{2} =
(\omega^{2} - 1) | f_{2}^{V} |^{2} + 2 (\omega^{2} - 1) | f^{*V} |^{2} \; .
\end{equation}

Assuming again pole behaviour of the helicity form factors we
obtain using Eq. (24) and the first relation in Eq.(22):
\begin{equation}
f_{2}^{V} = - f_{2}^{*A} = - f^{*V} \; .
\end{equation}

Eqs. (22) and (25) can be seen to be the HQET result Eq. (13) with $f_1^V=
F_1$ and $f_2^V=-F_2$. They do not represent a
unique solution and strictly speaking, we can only claim that the
HQL spectral function identities Eqs.(9) and (10) are consistent with the
explicit
form factor results following from the HQL. Imposing pole dominance of
the form factors, however, the HQL relations follow.

Finally, we consider the $B\rho$ and $B^{*} \rho$ intermediate states. The
helicity amplitudes for the $B\rho$ case are defined in analogy to Eq. (18)
. The helicity amplitudes for $B^*\rho^*$ are defined by
\begin{equation}
H_{\lambda_{W}; \lambda_{B^{*}}, \lambda_{\rho}} =
\langle 0 | J_{\mu} |B^{*} (\lambda_{B^{*}}) \rho (\lambda_{\rho}) \rangle
\epsilon^{\mu}(\lambda_{W})
\end{equation}
with $\lambda_{W} = \lambda_{B^{*}} - \lambda_{\rho}$ $\;$ . The helicity
amplitude expressions can be worked out but are too lenghty to be
reproduced here.

The equality $\rho_{VV}^{(0)} = \rho_{AA}^{(1)}$ for the $B \rho$ and
$B^{*} \rho$ system then implies
\begin{eqnarray}
3|H_{t;00}^{V} (B^{*} \rho) |^{2} + 6 |H_{t;++}^{V}(B^{*} \rho)|^{2}
         &=& 2|H_{+;0-}^{A} (B \rho) |^{2} + 2 | H_{+;+0}^{A} (B^{*} \rho)
         |^{2} + 2|H_{+;0-}^{A} (B^{*} \rho)|^{2} \nonumber \\[1ex]
& & + | H_{0;00}^{A} (B \rho)|^{2} + 2|H_{0;++}^{A} (B^{*} \rho) |^{2}
\end{eqnarray}
Similarly, it follows from $\rho_{AA}^{(0)} = \rho_{VV}^{(1)}$
that
\begin{eqnarray}
3|H_{t;00}^{A} (B \rho) |^{2} + 6 |H_{t;++}^{A}(B^{*} \rho)|^{2}
&=& 2|H_{+;0-}^{V} (B \rho) |^{2} + 2 | H_{+;+0}^{V} (B^{*} \rho)
      |^{2} +2|H_{+;0-}^{V} (B^{*} \rho)|^{2} \nonumber \\[1ex]
& &   + | H_{0;00}^{V} (B^{*} \rho)|^{2} + 2|H_{0;++}^{V} (B^{*} \rho) |^{2}
\end{eqnarray}
The HQL results in Eq.(13) can be seen to satisfy Eqs. (27) and (28).
It seems to be impossible, however, to derive the HQL results from these two
equations as they involve altogether ten form factors. All we can say is
that the spectral function identities are consistent with the HQL form factors.

In conclusion we have derived relations among form factors describing the
current-induced transitions: (vacuum) $\rightarrow B, B^{*}, B \pi, B^{*} \pi,
B \rho$ and $B^{*} \rho$ in the HQET. We show that many of these form factor
relations may be obtained in a simple manner by exploiting the identities
between the scalar and axial vector, and pseudoscalar and vector spectral
functions which hold in the heavy quark limit by considering in turn the
various one- and two-particle intermediate states.

\subsection*{\bf Acknowledgement}

Part of this work was done while two of us (JGK and KS) were visitors at
the University of Cape Town, Republic of South Africa. JGK and KS would like
to thank the Institute of Theoretical Physics at UCT for their hospitality.

\end{document}